\documentstyle[12pt]{article}

\title{METHODS OF MINIMIZATION OF CALCULATIONS IN HIGH ENERGY
       PHYSICS: \\
       {\sf II.}~Minimization of Number of Vectors in Problem}

\author {Alexander~L.~Bondarev
\and \it National Scientific and Educational Center of Particle and
\and \it High Energy Physics attached to Belarusian State University
\and \it M.Bogdanovich str.,153, Minsk 220040, Republic of Belarus
\and \rm e-mail: bondarev@hep.by}

\begin{document}
\maketitle

\begin{abstract}

    A number of different ways  of reducing the number of  vectors
describing  the  condition  of  particles  for high energy physics
problems are presented.   In particular  the decomposition of  any
vector with respect to the basis, consisting of any four  linearly
independent  vectors,  including  the  orthonormal  basis  and the
construction of orthonormal  bases from the  vectors of a  problem
(these  bases  may  be  used  as  the polarization ones for vector
bosons) as  well as  the expression  of one  vector of the problem
through the other are considered.

\end{abstract}


\section {Introduction}

    As is well known,  the maximum number of  linearly independent
vectors is equal  to four in  the Minkowski space.   Therefore for
any problem  the number  of vectors,  describing the  condition of
particles, may be reduced to four, by choosing four vectors as the
basic ones and by expressing all the other vectors through them.

    In Section {\bf 2} we  show, how to decompose any  vector with
respect to the basis, consisting of any four linearly  independent
vectors, including the case of the orthonormal basis.

    In Section {\bf 3} the construction of orthonormal bases from
the vectors of a problem is considered. In particular these bases
may be used as the polarization ones for vector bosons.

    In Section  {\bf 4}  some possibilities  of expression of some
vectors of a problem through the other are considered.

    The   methods   described   are   especially  effective  in  a
combination with the covariant  method for the calculating  of the
amplitudes  (see~\cite{r1},  \cite{r5},  \cite{r6},  \cite{r8}  --
\cite{r14}, \cite{r17}, \cite{r19},  \cite{r21}), but can  also be
used irrespective of it (see~\cite{r7}, \cite{r15}, \cite{r16}).


\section {The decomposition of any vector with respect to the
basis in the Minkowski space}

   Let us consider the determinant of the fifth order in
the Minkowski space
\begin {equation}
\begin {array}{l} \displaystyle
\left| \matrix{
g_{\mu \nu}   & g_{\mu \alpha}  &
g_{\mu \beta} & g_{\mu \lambda} &
g_{\mu \rho}  \\
g_{\sigma \nu}   & g_{\sigma \alpha}  &
g_{\sigma \beta} & g_{\sigma \lambda} &
g_{\sigma \rho}  \\
g_{\tau \nu}   & g_{\tau \alpha}  &
g_{\tau \beta} & g_{\tau \lambda} &
g_{\tau \rho}  \\
g_{\kappa \nu}   & g_{\kappa \alpha}  &
g_{\kappa \beta} & g_{\kappa \lambda} &
g_{\kappa \rho}  \\
g_{\omega \nu}   & g_{\omega \alpha}  &
g_{\omega \beta} & g_{\omega \lambda} &
g_{\omega \rho}
} \right| \equiv 0
\;
\end {array}
\label{e2.1}
\end {equation}
where
$$
\displaystyle
g_{\mu \nu}
= \left\{
\begin{array}{rl}
               1 & $if$ \;\;\; \mu = \nu = 0  \\
              -1 & $if$ \;\;\; \mu = \nu = 1,2,3  \\
               0 & $if$ \;\;\; \mu \neq \nu
\end{array}
\right.
$$

    The validity  of the  equation (\ref{e2.1})  in the  Minkowski
space   follows   from   the   properties   of   determinants  and
four-dimensionality  of   the  Minkowski   space  (see~\cite{r2}).
Really, using the properties  of determinants we notice,  that the
tensor  in  the  left-hand  side  of  the equation (\ref{e2.1}) is
completely antisymmetric  with respect  to each  of five  indices:
${\nu}$,   ${\alpha}$,   ${\beta}$,   ${\lambda}$,  ${\rho}$  (and
${\mu}$, ${\sigma}$,  ${\tau}$, ${\kappa}$,  ${\omega}$ as  well).
In a four-dimensional space,  every tensor which is  antisymmetric
with respect to more than four indices is equal to zero, since the
values of at least two of them must be equal.

   By the analogy with the Gram determinant (see~\cite{r3}) we
introduce the notation
$$
\begin {array}{l} \displaystyle
\left| \matrix{
g_{\mu \nu} & g_{\mu \alpha} & g_{\mu \beta} & g_{\mu \lambda} &
g_{\mu \rho}  \\
g_{\sigma \nu} & g_{\sigma \alpha} & g_{\sigma \beta} &
g_{\sigma\lambda} & g_{\sigma \rho}  \\
g_{\tau \nu} & g_{\tau \alpha} & g_{\tau \beta} & g_{\tau \lambda} &
g_{\tau \rho}  \\
g_{\kappa \nu} & g_{\kappa \alpha} & g_{\kappa \beta} &
g_{\kappa \lambda} &
g_{\kappa \rho}  \\
g_{\omega \nu} & g_{\omega \alpha} & g_{\omega \beta} &
g_{\omega \lambda} &
g_{\omega \rho}
} \right|
= G\pmatrix{\mu & \sigma & \tau  & \kappa  & \omega \\
            \nu & \alpha & \beta & \lambda & \rho }
\; .
\end {array}
$$

Let us consider%
\footnote{
We use the same metric as in the book \cite{r4}: \\
$
\displaystyle
a^{\mu} = ( a_0 , \vec{a} ) ,
\;\;\;
a_{\mu} = ( a_0, -\vec{a} ) ,
\;\;\;
ab = a_{\mu} b^{\mu} = a_0 b_0 - \vec{a} \vec{b}
\, $,
sign of the Levi-Civita tensor is determined as
$
\displaystyle
{\varepsilon}_{ 0 1 2 3 } = + 1
$
}

\begin {equation}
\begin {array}{l} \displaystyle
G\pmatrix{\mu & \sigma & \tau  & \kappa  & \omega \\
          \nu & \alpha & \beta & \lambda & \rho }
(l_0)^{\mu} (l_1)^{\sigma} (l_2)^{\tau}  (l_3)^{\kappa}  a^{\omega}
(l_0)^{\nu} (l_1)^{\alpha} (l_2)^{\beta} (l_3)^{\lambda}
        \\[0.5cm] \displaystyle
= G\pmatrix{l_0 & l_1 & l_2 & l_3 & a \\
            l_0 & l_1 & l_2 & l_3 & \rho }
= \left| \matrix{
{l_0}^2  & (l_0l_1) & (l_0l_2) & (l_0l_3) & (l_0)_{\rho} \\
(l_0l_1) & {l_1}^2  & (l_1l_2) & (l_1l_3) & (l_1)_{\rho} \\
(l_0l_2) & (l_1l_2) & {l_2}^2  & (l_2l_3) & (l_2)_{\rho} \\
(l_0l_3) & (l_1l_3) & (l_2l_3) & {l_3}^2  & (l_3)_{\rho} \\
(al_0)   & (al_1)   & (al_2)   & (al_3)   & a_{\rho}
} \right| = 0
\;  .
\end {array}
\label{e2.2}
\end {equation}

     From (\ref{e2.2}) we have
\begin {equation}
\begin {array}{l} \displaystyle
a = { 1 \over
G\pmatrix{l_0 & l_1 & l_2 & l_3 \\
          l_0 & l_1 & l_2 & l_3  } }
\Biggl[
    G\pmatrix{a   & l_1 & l_2 & l_3  \\
              l_0 & l_1 & l_2 & l_3   } l_0
  + G\pmatrix{l_0 & a   & l_2 & l_3  \\
              l_0 & l_1 & l_2 & l_3   } l_1
    \\    \\[0.5cm] \displaystyle
  + G\pmatrix{l_0 & l_1 & a   & l_3  \\
              l_0 & l_1 & l_2 & l_3   } l_2
  + G\pmatrix{l_0 & l_1 & l_2 & a    \\
              l_0 & l_1 & l_2 & l_3   } l_3
\Biggr]
\end {array}
\label{e2.3}
\end {equation}
where $G$ are the usual Gram determinants of the fourth order.

    If the condition
$$
\begin {array}{l} \displaystyle
G\pmatrix{l_0 & l_1 & l_2 & l_3 \\
          l_0 & l_1 & l_2 & l_3  }  \ne 0
\end {array}
$$
    is carried out,  that is if  the vectors $l_0$,  $l_1$, $l_2$,
$l_3$  are  linearly  independent,  then the equality (\ref{e2.3})
allows us  to decompose  any vector  $a$ with  respect to the four
vectors $l_0$, $l_1$, $l_2$, $l_3$.

    If the vectors $l_0$, $l_1$, $l_2$, $l_3$ form the orthonormal
basis, that is if
$$
\displaystyle
(l_{\mu}l_{\nu}) =  g_{\mu \nu}
$$
then (\ref{e2.2}) takes the form
\begin {equation}
\begin {array}{l} \displaystyle
\left| \matrix{
1       & 0       & 0       & 0       & (l_0)_{\rho} \\
0       & -1      & 0       & 0       & (l_1)_{\rho} \\
0       & 0       & -1      & 0       & (l_2)_{\rho} \\
0       & 0       & 0       & -1      & (l_3)_{\rho} \\
(al_0)  & (al_1)  & (al_2)  & (al_3)  & a_{\rho}
} \right| = 0
\;  .
\end {array}
\label{e2.4}
\end {equation}

    And it follows that
\begin {equation}
\displaystyle
a = (a l_0) l_0 - (a l_1) l_1 - (a l_2) l_2 - (a l_3) l_3
\; .
\label{e2.5}
\end {equation}


\section {The construction of orthonormal  bases from the vectors
of a problem}

    With the  help of  three vectors  of a  problem and completely
antisymmetric  Levi-Civita  tensor  we  can  always  construct  an
orthonormal basis.  Let us consider the three bases of this sort.

\begin{enumerate}
\item
Let $p$ be an arbitrary $4$-momentum such that
$$
\displaystyle
p^2 = m^2 \ne 0
\; ,
$$
    $a$ and $b$  are arbitrary vectors.   Then the  following four
vectors $l_0$, $l_1$, $l_2$, $l_3$ form the orthonormal basis:
\begin {equation}
\displaystyle
l_0 = { p \over m }
\; ,
\label{e3.1}
\end {equation}
\begin {equation}
\begin {array}{l} \displaystyle
(l_1)_{\rho} = - { G\pmatrix{p & \rho \\
                             p & a   }
 \over  m
\Biggl[ -
 G\pmatrix{p & a  \\
           p & a   }
\Biggr]^{1/2} } =
{ (pa) p_{\rho} - m^2 a_{\rho} \over
m \sqrt{ (p a)^2 - m^2 a^2 }  }
\; ,
\end {array}
\label{e3.2}
\end {equation}
\begin {equation}
\begin {array}{l} \displaystyle
(l_2)_{\rho} = { G\pmatrix{p & a & \rho \\
                           p & a & b  }
 \over
\Biggl[ -
 G\pmatrix{p & a  \\
           p & a   }
 G\pmatrix{p & a & b  \\
           p & a & b   }
\Biggr]^{1/2} }
        \\[0.5cm] \displaystyle
= { \left[ (p a) (a b) - a^2 (p b) \right] p_{\rho}
  + \left[ (p a) (p b) - m^2 (a b) \right] a_{\rho}
  + \left[ m^2 a^2 - (p a)^2 \right] b_{\rho}
\over
 \sqrt{ (p a)^2 - m^2 a^2 }
 \sqrt{ 2(p a) (p b) (a b) + m^2 a^2 b^2 -
 m^2 (a b)^2 - a^2 (p b)^2 - b^2 (p a)^2 } }
\; ,
\end {array}
\label{e3.3}
\end {equation}
\begin {equation}
\begin {array}{l} \displaystyle
(l_3)_{\rho}
= { {\varepsilon}_{ {\rho} {\alpha} {\beta} {\lambda} }
  p^{\alpha} a^{\beta} b^{\lambda}
       \over
\Biggl[
 G\pmatrix{p & a & b  \\
           p & a & b   }
\Biggr]^{1/2} }
        \\[0.5cm] \displaystyle
= { {\varepsilon}_{ {\rho} {\alpha} {\beta} {\lambda} }
p^{\alpha} a^{\beta} b^{\lambda}
\over
 \sqrt{ 2(p a) (p b) (a b) + m^2 a^2 b^2 -
 m^2 (a b)^2 - a^2 (p b)^2 - b^2 (p a)^2 } }
\; .
\end {array}
\label{e3.4}
\end {equation}

     In particular, the authors of \cite{r5}, \cite{r6} use
the following set:
$$
\displaystyle
p = p_i
\, , \;
a = p_f
\, , \;
b = r
$$
where
$
\displaystyle
p_i
$
and
$
\displaystyle
p_f
$
are the $4$-momenta  of the initial and final particles
respectively and $r$ is an arbitrary  $4$-momentum.

   In \cite{r7} for the construction of the basis the following
vectors are used:
$$
\displaystyle
p = p_i
\, , \;
a = q_i
\, , \;
b = q_f
$$
where
$
\displaystyle
q_i
$
and
$
\displaystyle
q_f
$
are the $4$-momenta of the initial and final particles of the
other line of the diagram respectively.

   In \cite{r8} the construction of the basis uses the following
vectors
$$
\displaystyle
p = q_i
\, , \;
a = q_f
\, , \;
b = p_i
$$
(in this paper only the last three vectors of the basis are used).

  In \cite{r9} -- \cite{r13} a special form of the basis
(\ref{e3.1}) -- (\ref{e3.4}) with
$$
\displaystyle
a^{\mu} = (m,0,0,0)
\, , \;
b^{\mu} = (0,0,0,1)
$$
is used.
(In \cite{r11} -- \cite{r13} only the last three vectors of
the basis are used, in \cite{r13} there is the restriction
$
\displaystyle
p_y = 0 \,
$.)

\item
     Let $p_1 \, , \; p_2$ are the $4$-momenta of a problem such
that
$$
\displaystyle
p_1^2 = m_1^2 \ne 0
\, , \;
p_2^2 = m_2^2  \ne 0
\, .
$$
Let us consider a special form of the basis
(\ref{e3.1})~--~(\ref{e3.4}) at
$$
p = { m_2 p_1 + m_1 p_2 \over \sqrt { m_1 m_2 } }
\, , \;
m = \sqrt { 2 \left[ ( p_1 p_2 ) + m_1 m_2 \right] }
\, , \;
a = p_2 \, :
$$
\begin {equation}
\displaystyle
l_0 = { m_2 p_1 + m_1 p_2 \over
\sqrt{ 2 m_1 m_2 \left[ (p_1 p_2) + m_1 m_2 \right] } }
\; ,
\label{e3.5}
\end {equation}
\begin {equation}
\displaystyle
l_1 = { m_2 p_1 - m_1 p_2 \over
\sqrt{ 2 m_1 m_2 \left[ (p_1 p_2) - m_1 m_2 \right] } }
\; ,
\label{e3.6}
\end {equation}
\begin {equation}
\begin {array}{l} \displaystyle
(l_2)_{\rho} = { G\pmatrix{p_1 & p_2 & \rho \\
                           p_1 & p_2 & b  }
 \over
\Biggl[ -
 G\pmatrix{p_1 & p_2  \\
           p_1 & p_2   }
 G\pmatrix{p_1 & p_2 & b  \\
           p_1 & p_2 & b   }
\Biggr]^{1/2} }
        \\[0.5cm] \displaystyle
= { \left[ (p_1 p_2) (p_2 b) - m_2^2 (p_1 b) \right]
(p_1)_{\rho} +
  \left[ (p_1 p_2) (p_1 b) - m_1^2 (p_2 b) \right]
(p_2)_{\rho} +
  \left[ m_1^2 m_2^2 - (p_1 p_2)^2 \right] b_{\rho}
\over
 \sqrt{ (p_1 p_2)^2 - m_1^2 m_2^2 }
 \sqrt{ 2(p_1 p_2) (p_1 b) (p_2 b) + m_1^2 m_2^2 b^2 -
 m_1^2 (p_2 b)^2 - m_2^2 (p_1 b)^2 - b^2 (p_1 p_2)^2 } }
\; ,
\end {array}
\label{e3.7}
\end {equation}
\begin {equation}
\begin {array}{l} \displaystyle
(l_3)_{\rho}
= { {\varepsilon}_{ {\rho} {\alpha} {\beta} {\lambda} }
   p_1^{\alpha} p_2^{\beta} b^{\lambda}
 \over
\Biggl[
 G\pmatrix{p_1 & p_2 & b  \\
           p_1 & p_2 & b   }
\Biggr]^{1/2} }
        \\[0.5cm] \displaystyle
= { {\varepsilon}_{ {\rho} {\alpha} {\beta} {\lambda} }
p_1^{\alpha} p_2^{\beta} b^{\lambda}
\over
 \sqrt{ 2(p_1 p_2) (p_1 b) (p_2 b) + m_1^2 m_2^2 b^2 -
 m_1^2 (p_2 b)^2 - m_2^2 (p_1 b)^2 - b^2 (p_1 p_2)^2 } }
\; .
\end {array}
\label{e3.8}
\end {equation}
    In \cite{r6} for the construction of the basis the vectors
$$
\displaystyle
p_1 = p_i
\, , \;
p_2 = p_f
\, , \;
b = r
$$
are used, where
$
\displaystyle
p_i
$
and
$
\displaystyle
p_f
$
are the $4$-momenta of the initial and final particles
respectively and $r$ is an arbitrary  $4$-momenta.

   In \cite{r14} for the construction of the basis the same
vectors are used under the additional condition
$$
\displaystyle
(p_i r) = (p_f r) = 0
\; .
$$

   In \cite{r15}  the basis is constructed with the help of
$$
\displaystyle
p_1 = p_f
\, , \;
p_2 = p_i
\, , \;
b = q_i
$$
or
$$
\displaystyle
p_1 = p_i
\, , \;
p_2 = p_f
\, , \;
b = q_i
$$
   (in this paper only vectors
$
l_2
$
and
$
l_3
$
are used).

\item
     Let $p_1 \, , \; p_2$ be the $4$-momenta of a problem such
that
$
\displaystyle
p_1^2 = p_2^2 = 0 \, .
$
Let us consider a special form of the basis
(\ref{e3.1})~--~(\ref{e3.4}) at
$$
p =  p_1 +  p_2
\, , \;
m = \sqrt { 2 ( p_1 p_2 ) }
\, , \;
a = p_2 \, :
$$
\begin {equation}
\displaystyle
l_0 = { p_1 + p_2 \over \sqrt{ 2 (p_1 p_2) } }
\; ,
\label{e3.9}
\end {equation}
\begin {equation}
\displaystyle
l_1 = { p_1 - p_2 \over \sqrt{ 2 (p_1 p_2) } }
\; ,
\label{e3.10}
\end {equation}
\begin {equation}
\displaystyle
(l_2)_{\rho}
= { (p_2 b) (p_1)_{\rho} + (p_1 b) (p_2)_{\rho}
  - (p_1 p_2) b_{\rho}
\over
 \sqrt{ 2(p_1 p_2) (p_1 b) (p_2 b) - b^2 (p_1 p_2)^2 } }
\; ,
\label{e3.11}
\end {equation}
\begin {equation}
\displaystyle
(l_3)_{\rho}
= { {\varepsilon}_{ {\rho} {\alpha} {\beta} {\lambda} }
     p_1^{\alpha} p_2^{\beta} b^{\lambda}
 \over
 \sqrt{ 2(p_1 p_2) (p_1 b) (p_2 b)  - b^2 (p_1 p_2)^2 } }
\; .
\label{e3.12}
\end {equation}

   In \cite{r16} the basis is constructed with the help of
the vectors
$$
\displaystyle
p_1 = k_1
\, , \;
p_2 = k_2
\, , \;
b = p_i - p_f
$$
where
$
k_1
$
and
$
k_2
$
are the $4$-momenta of photons.

    In \cite{r17}  for the construction of the basis the vectors
$$
\displaystyle
p_1 = k
\, , \;
p_2 = p_i
\, , \;
b = p_f
$$
or
$$
\displaystyle
p_1 = k
\, , \;
p_2 = p_f
\, , \;
b = p_i
$$
are used, where $k$ is the $4$-momentum of photon
(in this work only the vectors
$
l_2
\, , \;
l_3
\,
$
are used).
\end{enumerate}

    An appropriate choice of the form of the basis and the vectors
of a  problem for  its construction  can essentially  simplify the
calculations.

Note that in \cite{r22} it was proposed  the decomposition of any
4-vector $q$
$$
\displaystyle
q^{\mu} = { 1 \over \varepsilon(p_1, p_2, p_3, p_4) }
\left[
(qp_1) v_1^{\mu} + (qp_2) v_2^{\mu} + (qp_3) v_3^{\mu}
+ (qp_4) v_4^{\mu}
\right]
$$
with respect to the basis
\begin {equation}
\begin {array}{l} \displaystyle
v_1^{\mu} =
\varepsilon^{\mu \nu \rho \lambda}
(p_2)_{\nu} (p_3)_{\rho} (p_4)_{\lambda}
\, , \;
v_2^{\mu} =
\varepsilon^{\nu \mu \rho \lambda}
(p_1)_{\nu} (p_3)_{\rho} (p_4)_{\lambda}
\, , \;
     \\     \displaystyle
v_3^{\mu} =
\varepsilon^{\nu \rho \mu \lambda}
(p_1)_{\nu} (p_2)_{\rho} (p_4)_{\lambda}
\, , \;
v_4^{\mu} =
\varepsilon^{\nu \rho \lambda \mu}
(p_1)_{\nu} (p_2)_{\rho} (p_3)_{\lambda}
\, ,
\end {array}
\label{e3.13}
\end {equation}
where
$
\displaystyle
\varepsilon(p_1, p_2, p_3, p_4)
= \varepsilon^{\mu \nu \rho \lambda}
(p_1)_{\mu} (p_2)_{\nu} (p_3)_{\rho} (p_4)_{\lambda}
\,  ;  \, \;
p_1 \, , \; p_2 \, , \; p_3 \, , \; p_4 \,
$
are arbitrary vectors of a problem.

    However the basis (\ref{e3.13})  is not orthonormal.   Besides
it  is  necessary  to  use  four  vectors  of  a  problem  for its
construction.


\section {The expression of vectors of a problem through
the other}

\begin{enumerate}
\item
    It is possible to express  the vector $n$ which determine  the
axis of the spin projections  of a fermion through the  4-momentum
of this fermion and any other vector.  Really, let us consider the
vector
\begin {equation}
\displaystyle
n(p,a) =  { (p a) p - m^2 a
 \over
  m \sqrt{ (p a)^2 - m^2 a^2 } }
\; .
\label{e4.1}
\end {equation}
where $a$ is an arbitrary vector. One can easily see that the vector
$
n(p,a)
$
satisfies the standard conditions
$$
\displaystyle
n^2 = -1
\, , \;
(p n) = 0
\, .
$$

    Choosing
$
\displaystyle
a^{\mu} = (1, \, 0,0,0) \, ,
$
we have
$$
\displaystyle
n^{\mu} = { 1 \over m }
(|\vec{p}| \, , \; { p_0 \over |\vec{p}| } \vec{p} )
$$
that is in this case the state of polarization of a particle is
the helicity (see e.g.~\cite{r18}).

  In \cite{r5} -- \cite{r6} it was proposed to choose
$$
\displaystyle
n_f =  { (p_i p_f) p_f - m_f^2  p_i
 \over
  m_f \sqrt{ (p_i p_f)^2 - m_i^2 m_f^2 } }
$$
that is in this case
$
\displaystyle
a = p_i
\, .
$

  In \cite{r7}
$
\displaystyle
a = q_i
\, , \;
$
where  $ q_i $ is the 4-momentum of the initial particle of
the other line of the diagram.

   In \cite{r19}
$
\displaystyle
a = q
\, , \;\;
$
where
$
\displaystyle
q^2 = 0
$
and for the numerical calculations there is taken the vector
$
\displaystyle
q^{\mu} = (1, \, 1,0,0) \, .
$

   In \cite{r20}
$
\displaystyle
a^{\mu}
= \left[ ( {\vec p} {\vec o} ) , ( p_0 + m ) {\vec o} \right]
\, , \;\;
$
where
$
\displaystyle
{\vec o}^2 = 1
$.

    A method of this sort suitable when we are not interested in
the polarization state of the particle and therefore have  to sum
or average over its polarizations.

\item
   In the paper \cite{r21} for an arbitrary vector $S$ such that
$$
\displaystyle
S^2 \ne 0
$$
there was offered the decomposition
\begin {equation}
\displaystyle
S = \left[ S - { S^2 \over 2(q S) } q \right] +
  { S^2 \over 2(q S) } q  = s +  { S^2 \over 2(q S) } q
\, ,
\label{e4.2}
\end {equation}
where $q$ is an arbitrary vector such that
$$
\displaystyle
q^2 = 0
\, .
$$
  As this takes place, the vector
$$
\displaystyle
s = S - { S^2 \over 2(q S) } q
$$
also has the property
$$
\displaystyle
s^2 = 0
\, .
$$

   This method is especially effective for the calculations for
the processes involving massless particles.
\end{enumerate}



\end {document}